\documentstyle[12pt,epsf]{lamuphys}
\makeatletter
\let\chapter\hid@chapter
\makeatother
\begin{document}
\input{psfig}
\pagenumbering{arabic}
\title{Stellar populations in bulges of spiral galaxies}

\author{Pascale Jablonka}

\institute{DAEC-URA173, Observatoire de Paris-Meudon, France }

\maketitle

\begin{abstract}
We discuss the integrated properties of the stellar population in bulges along
the Hubble sequence and new HST data for individual stars
in the bulge of M31.
\end{abstract}
\section{Bulges along the Hubble sequence}

Bulge stellar populations are still poorly investigated.  It remains
unclear whether bulges are formed in the very early stages of galaxy
formation in synchronicity with the halo, or on the contrary on longer
time scales and from disk material. The effective role of bars in
bulge growth and star formation is also to be clarified. The
bulge-to-disk light ratio varies along the Hubble sequence, and it is
important to understand whether this trend translates into a mass
scale only or into differences of star formation history as well.  For a
long time, the bulge of our Galaxy has been considered as the prototype of
bulges in general. However, no blame on this simplification, since the
study of bulges, although imperative, has to cope with the difficult
problem of the contamination of bulge by disk light.

\begin{figure}[h]
\centerline{
\psfig{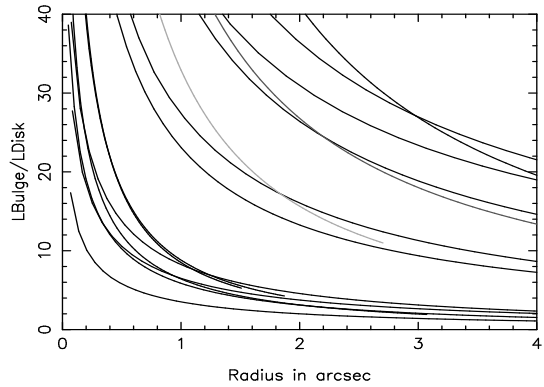}}
\caption{The radial variation of the bulge-to-disk luminosity ratio
in 28 spiral galaxies.}
\end{figure}

Taking advantage of Kent (1985) work on decomposition between disk and
bulge lights, we show in Fig.~1, for a few face-on spiral galaxies,
the radial variation of their bulge-to-disk light ratio,
L$_{Bulge}$/L$_{Disk}$. It is conspicuous that the integration of
light in a constant aperture for all these galaxies would gather
different amount of disk light. This would severely affect any search
for trends of bulge properties along the Hubble sequence. Also, this
disk light contamination can be present even in the central regions of
galaxies.  Therefore, observations of bulges require the acceptance of
some disk light contamination, yet this contamination must be kept as
small as possible and in equal proportion for all galaxies; this can be
done by having a different aperture for each galaxy.

The integrated-light spectra of 28 spiral galaxies were obtained in 1993,
using an automatic drift scanning procedure available at the Steward
Observatory 2.3-m telescope equipped with a Boller \& Chivens CCD
spectrograph. For each galaxy, structural parameters, such as
effective radius $r_e$, scale height $h$, surface brightness values
$\mu_e$ at $r_e$ and $\mu_0$ at the center were available (Kent 1985).
Consequently, we were able to calculate the radius corresponding to a
fixed value of the bulge-to-disk luminosity ratio, a ratio equal to 6
for all sample galaxies.

\begin{figure}[h]
\centerline{
\psfig{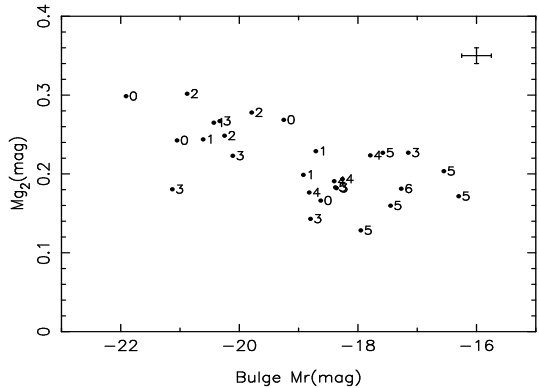}}
\caption{ The relation between the bulge total r-Gunn luminosity and
Mg$_2$ index.  Numbers represent the galaxies Hubble types (T
parameter). Typical maximum errors are shown.}
\end{figure}

\begin{figure}[h]
\centerline{
\psfig{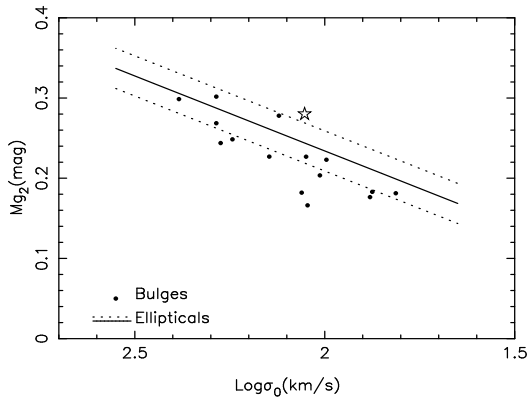}}
\caption{ The variation of the bulge Mg$_2$ index with the central
velocity dispersion $\sigma_0$. Plain and doted lines represent the
mean relation and 1-$\sigma$ dispersion known for ellipticals (Gorgas
\& Gonz\`alez, private communication). The open asterisk indicates the
bulge of our Galaxy.}
\end{figure}

Fig.~2 presents the variation of the bulge Mg$_2$ index with its
absolute magnitude in r-Gunn photometric band. The Hubble type of each
galaxy is indicated by its corresponding number. There is clearly
delineated a metallicity-luminosity relation for bulges along the
Hubble sequence: the bulge of our Galaxy can no longer be considered
as a template for all bulges. If bulges tend to be fainter in
late-type spirals than in early-type ones, this is however the only
segregation seen in Fig.~2.  Besides this tendency, all Hubble types
are mixed; the bulge luminosity is certainly the dominant factor in the
correlation.

Fig.~3 displays the relation existing between the bulge Mg$_2$ index
and the galaxy central velocity dispersion, when this information is
available. This is the case for about half the galaxies in our
sample. Our own Galaxy is indicated by an asterisk and the mean
relation obtained for elliptical galaxies (Gorgas \& Gonz\`alez,
private communication) is superimposed with its 1-$\sigma$
dispersion. This figure underlines how bulges and elliptical galaxies
are closely related systems and suggest common processes of
formation.

\section{The bulge of M31}

The high inclination between the disk of M31 and its line-of-sight
allows the study of the stellar populations of its bulge. The view we
have of it is global and essentially free of pollution by disk stars,
unlike the case for the bulge of our Galaxy.

We obtained, with the WFPC2 camera on the Hubble Space Telescope,
Cycle~5 and ~6 high spatial resolution images, in filters F555W and
F814W, of a few fields centered on super-metal rich star clusters in
the bulge of M31, clusters for which we have ground-based
spectrophotometric data (Jablonka et al. 1992).  Two clusters, G170
and G177, are located SW along the major axis of M31, respectively at
6.1 and 3.2 arcmin of the galaxy nucleus; another cluster, G198, is
located NE along the major axis at 3.7 arcmin (Huchra et al. 1991).
Adopting 1 arcmin = 250~pc from Rich \& Mighell (1995), these
separations correspond to projected distances of about 1.55, 0.80, and
0.92 kpc, respectively.  In addition to the cluster stellar
populations, these HST data give us the opportunity to study the
stellar populations in the surrounding bulge fields.  We present
hereafter some of the results on the latter point, which are part of
an extensive work to be published elsewhere (Jablonka et al.  1997).

Fig.~4 displays the cluster G177 and its surrounding bulge stellar
field, as observed in the PC frame.  The field is 36 arcsec by 36
arcsec in size. It is a composite image from the F814W and F555W
frames.  The total exposure time is 6500s in the filter F555W and
6000s in the filter F814W.  This figure illustrates the compactness of
the cluster G177 and the richness of its surrounding field.

\begin{figure}[h]
\centerline{
\psfig{bbllx=5mm,bblly=10mm,bburx=200mm,bbury=259mm,%
figure=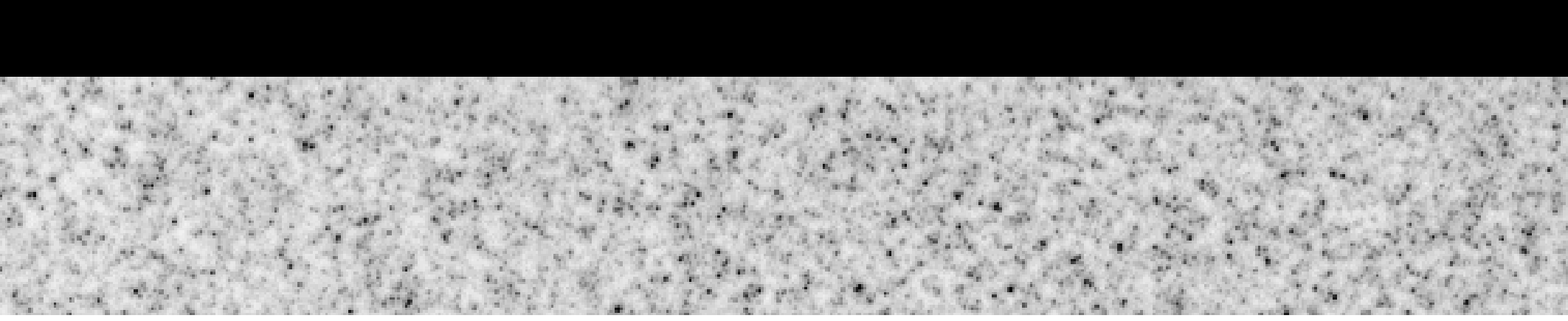,width=50mm,height=50mm}}
\caption{The 36$^{\prime\prime}$ $\times$ 36$^{\prime\prime}$ PC image centered on the
cluster G177}
\end{figure}

We used the DAOPHOT/ALLSTAR/ALLFRAME software package for
crowded-field photometry (Stetson 1994), along with the PSFs kindly
provided to us by the Cepheid-Distance-Scale HST key project. The
F555W and F814W instrumental magnitudes are {\it in fine} converted to
Johnson V and Cousins I magnitudes. Fig.~5 displays the resulting
(I,V--I) color-magnitude diagram obtained when the entire PC field
around G177 is analyzed. About 65,000 stars are detected.

\begin{figure}[h]
\centerline{
\psfig{bbllx=30mm,bblly=72mm,bburx=183mm,bbury=227mm,%
figure=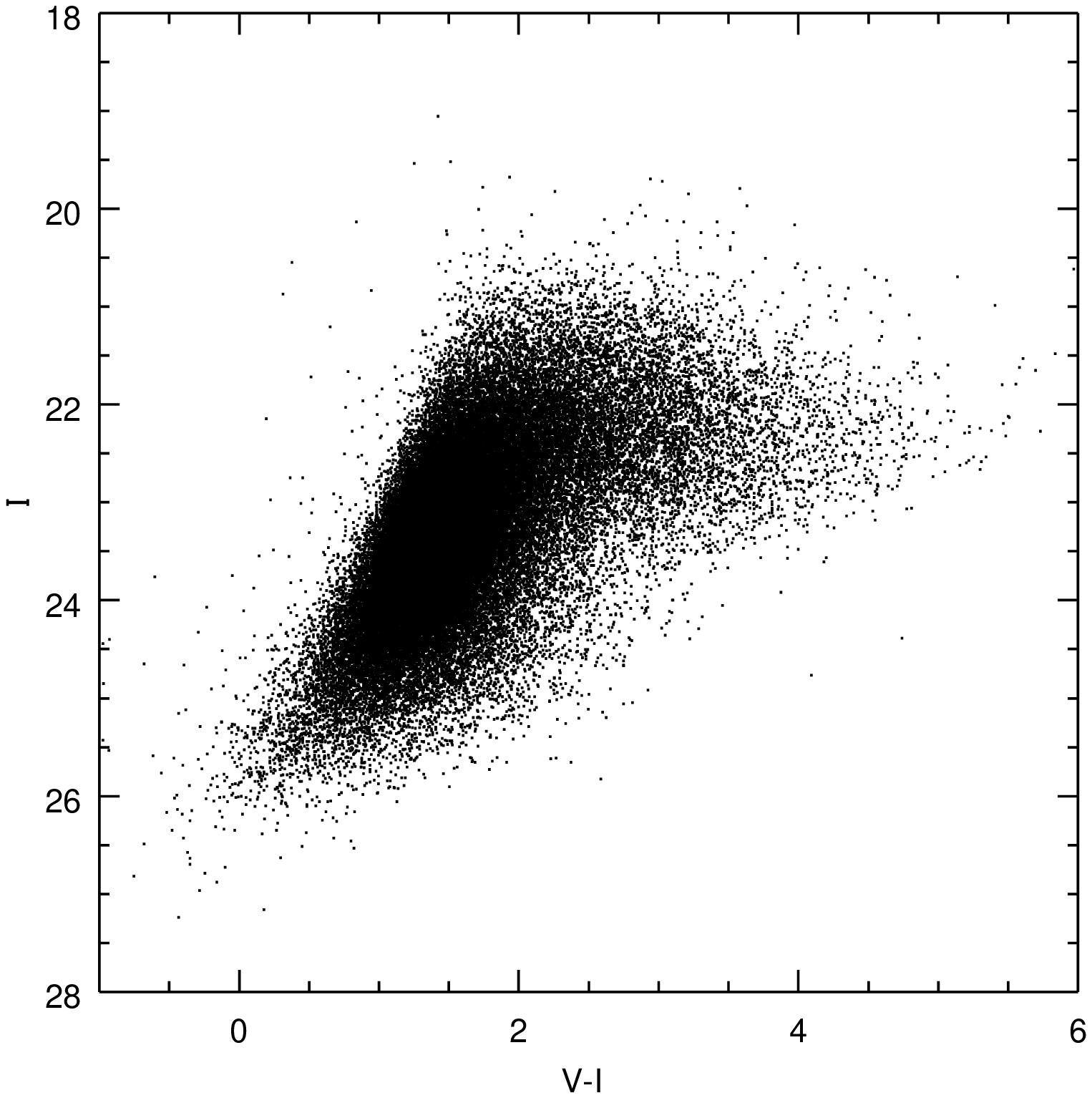,width=70mm,height=50mm}}
\caption{The color-magnitude diagram for 65,000 stars in the field
around the cluster G177, located 3.2 arcmin SW of M31 nucleus}
\end{figure}

The very high density of stars in this field appears to be the main
limiting factor in the detection of faint-magnitude stars, a problem
which a longer exposure time might not straightforwardly solve.  Our
photometry reaches the horizontal-branch (HB) level, but prevents us
to see its morphology.  The red-giant branch (RGB) is however clearly
observed, indicating a very metal-rich old population.

We are not the first ones to be attracted by the rich mine of
information provided by M31.  In particular, Mould (1986), Mould \&
Kristian (1986), and Rich \& Mighell (1995) gathered ground-based and
WFPC (aberrated HST) data for various stellar fields located between
0.2 and 12~kpc from the galaxy center.

Rich \& Mighell (1995), synthesizing their own and previous works,
discuss an apparent brightening of the RGB tip, when moving along a
sequence of stellar fields at distances decreasing from the halo to
the galaxy center: at 7~kpc, the RGB tip of stars in the field appears
at the same luminosity as the one observed for globular clusters (I
$\sim$ 20.5~mag), while it becomes 1~mag brighter when closer to the
M31 center (I $\sim$ 19.5~mag).  Rich \& Mighell (1995) mention the
contradiction between these observations and what can be expected from
a luminosity variation due to metallicity, since metal-rich globular
clusters have fainter RGBs (Bica et al. 1991).  They mention the
hypotheses of an M31 bulge younger by 5 to 7~Gyr than the extreme
Galactic halo and/or of the presence of a rare stellar evolutionary
phase.

However, we do not confirm the detection of these very luminous stars,
and our data are in perfect agreement with an increase of metallicity
towards the center of the galaxy.  The refurbished Hubble Space
Telescope brings a new light on this problem: 65,000 stars detected in
a 36$^{\prime\prime}$ $\times$ 36$^{\prime\prime}$ area, is equivalent
to about 50 stars per 1 arcsec$^2$. Only the refurbished HST with its
high spatial resolution capability is able to resolve these stars. It
seems very possible that previous studies suffered from crowding
problems and detected blends instead of individual stars, thus
artificially overestimating their luminosities.

The bulge stellar populations around the two other M31 clusters G198
and G170 present the same characteristics of those observed for G177
and shown in Fig.~5.  At this stage of the analysis, it is already
clear that M31 bulge stars share the same mean locus in their
color-magnitude diagrams as the two Galactic globular clusters
NGC~6528 and NGC~6553 (Ortolani et al. 1995).  These results favor a
rapid formation of bulges during the early history of galaxies.

\section{Acknowledgements}
I wish to thank the conference organizers for such an interesting and
lively meeting. I also wish to thank my collaborators, T. Bridges, A.
Maeder, G. Meylan, G. Meynet, A. Sarajedini, for allowing me to quote
some preliminary results before publication.

%
%
%

\end{document}